\RequirePackage{fix-cm}
\documentclass[twocolumn,epjc3]{svjour3}  
\smartqed  % flush right qed marks, e.g. at end of proof
\RequirePackage{graphicx}
 \RequirePackage{mathptmx}      % use Times fonts if available on your TeX system
\RequirePackage{latexsym}
\RequirePackage[numbers,sort&compress]{natbib}
\RequirePackage[colorlinks,citecolor=blue,urlcolor=blue,linkcolor=blue]{hyperref}
\RequirePackage{amsmath}
\journalname{Eur. Phys. J. A}
\begin{document}
\sloppy 
\title{Correspondence of multiplicity and energy distributions%\thanksref{t1}
}

%\titlerunning{Correspondence of multiplicity and energy distributions}        % if too long for running head

\author{Maciej Rybczy\'{n}ski\thanksref{e1,addr1} %etc.
        \and
        Zbigniew W\l odarczyk\thanksref{addr1} %etc.
}

\thankstext{e1}{e-mail: maciej.rybczynski@ujk.edu.pl}


\institute{Institute of Physics, Jan Kochanowski University, 25-406 Kielce, Poland \label{addr1}
%           Second address \label{addr2}
%           \emph{Present Address:} if needed\label{addr3}
}

\date{Received: date / Accepted: date}

\maketitle

\begin{abstract} 
The evaluation of the number of ways we can distribute energy among a collection of particles in a system is important in many branches of modern science. In particular, in multiparticle production processes the measurements of particle yields and kinematic distributions are essential for characterizing their global properties and to develop an understanding of the mechanism for particle production. We demonstrate that energy distributions are connected with multiplicity distributions by their generating functions.

%\PACS{02.50.Ey, 05.10.Ln, 12.40.Ee}
\end{abstract}

%\section{Introduction}

For the count probability distribution, $P\left(N\right)$, the generating function $G\left(z\right)$ is defined as:
\begin{equation}
G\left(z\right)=\sum_{N=0}^{\infty}P\left(N\right)z^{N}.
\end{equation}
Thus far the dummy variable $z$ of the generating function has been considered just as a technical auxiliary variable ("book keeping variable"). Only in the so called method of \textit{collective marks} one gives a probability interpretation for the variable $z$~\footnote{The method of collective marks was originated by van Dantzig~\cite{VanDantzig}, and discussed in~\cite{Runnenburg} and~\cite{Kleinrock}. Recently, the collective marks method was used to find the probability generating function for first passage probabilities of Markov chains~\cite{Zhang}.}. If we mark each of the $N$ elements in the set independently with probability $1-z$ and leave it unmarked with probability $z$, then $G\left(z\right)$ is the probability that there is no mark in the whole set.

In this letter multiplicity distributions $P\left(N\right)$ in quasi power-law ensembles and their generating functions $G\left(z\right)$ are discussed. They are connected with the energy distributions $F\left(E\right)$ of elements in the ensemble.

\begin{table}[ht!]
\centering
\caption{Distributions $P\left(N\right)$ used in this work: Poisson (PD), Negative Binomial (NBD) and Binomial (BD) and their generating functions $G\left(z\right)$.} 
\label{tab-1}
\begin{tabular*}{\columnwidth}{@{\extracolsep{\fill}}ccc@{}}
%\begin{tabular}{c|c|c}
\hline
 & $P\left(N\right)$ &  $G\left(z\right)$\\
\hline
PD & $\frac{\lambda^{N}}{N!}\exp\left(-\lambda\right)$ & $\exp\left[\lambda\left(z-1\right)\right]$ \\
%\hline
NBD  & $\frac{\Gamma\left(N+k\right)}{\Gamma\left(N+1\right)\Gamma\left(k\right)}p^{N}\left(1-p\right)^{k}$ & $\left[1-\frac{p}{1-p}\left(z-1\right)\right]^{-k}$ \\
%\hline
BD & $\frac{K!}{N!\left(K-N\right)!}p^{N}\left(1-p\right)^{K-N}$ & $\left[1+p\left(z-1\right)\right]^{K}$ \\
\hline
\end{tabular*}
\end{table}

Note, that generating functions of NBD and BD (shown in Table 1) are in fact some quasi-power functions of $z$ and as such can be written in the form of the corresponding Tsallis distributions~\cite{Tsallis:1987eu,Tsallis:2008mc,Tsallisbook,Wilk:2014zka}.
\begin{align}
G\left(z\right)&=\exp_{q}\left[\langle N\rangle\left(1-z\right)\right]\nonumber\\
               &=\left[1+\left(q-1\right)\langle N\rangle\left(1-z\right)\right]^{\frac{1}{1-q}},
\label{eq:gen}
\end{align}
where $q-1=1/k$ for NBD, $q-1=-1/K$ for BD, and $q-1\rightarrow 0$ for PD.
For 
\begin{equation}
z=1-\frac{E}{U}
\end{equation}
with the total available energy
\begin{equation}
U=\sum_{i=1}^{N} E_{i},
\end{equation}
the multiplicity generating function (\ref{eq:gen}) gives the energy distribution
\begin{equation}
F\left(E\right)=G\left(z=1-E/U\right)=\left[1+\left(q-1\right)\frac{E}{T}\right]^{\frac{1}{1-q}}
\label{hag-form}
\end{equation}
which is the well known Tsallis distribution~\cite{Tsallis:1987eu}, and which for $q\rightarrow 1$ becomes Boltzmann-Gibbs distribution. This distribution was first proposed in~\cite{Michael:1976pz,Michael:1977hx} as the simplest formula extrapolating exponential behavior observed for low transverse momenta to power law behavior at large transverse momenta. At present it is known as the QCD-inspired \textit{Hagedorn formula}~\cite{Arnison:1982ed,Hagedorn:1983wk}. Function (\ref{hag-form}) is usually interpreted in terms of the statistical model of particle production employing the Tsallis non-extensive statistics~\cite{Tsallis:1987eu,Tsallis:2008mc,Tsallisbook} and widely used in description of multiparticle production processes~\cite{Wong:2015mba,Wilk:2012zn}~\footnote{For an  updated  bibliography  on  this  subject,  see {\tt http://tsallis.cat.cbpf.br/biblio.htm}}. 
\begin{figure*}
\centering
\includegraphics[angle=0,width=0.49 \textwidth]{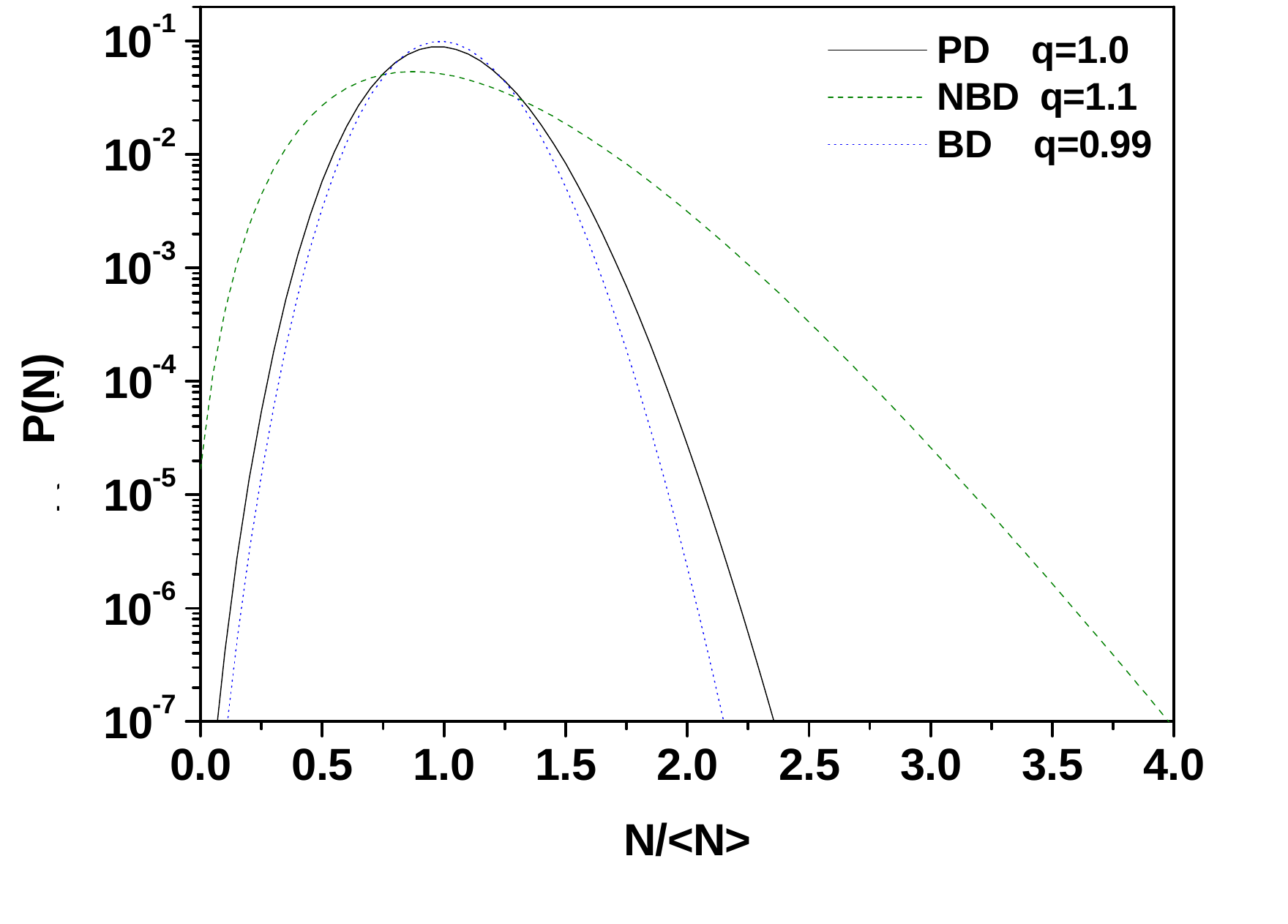} 
\includegraphics[angle=0,width=0.49 \textwidth]{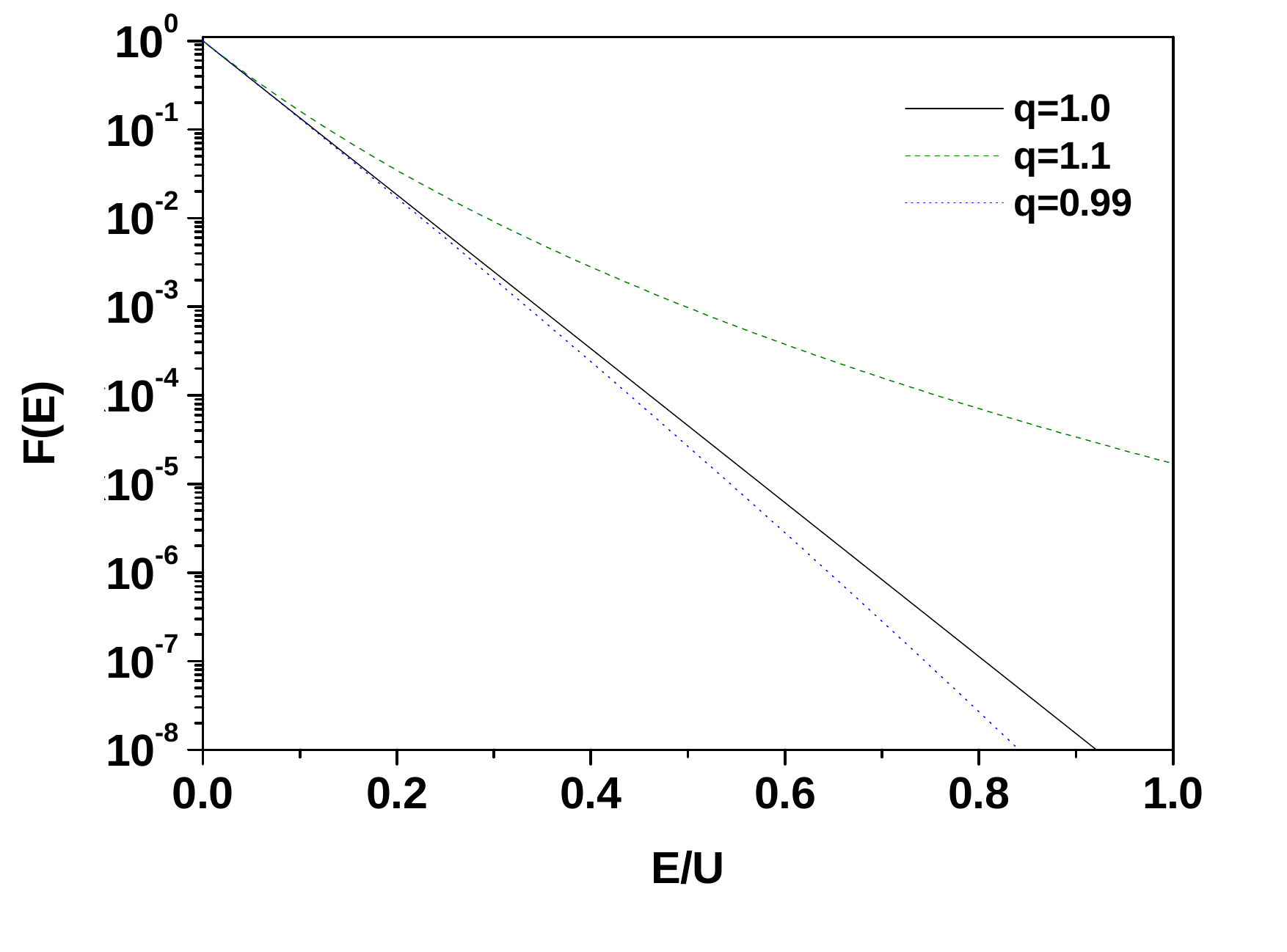} 
\vspace{-4mm}
\caption {Multiplicity distributions and corresponding energy distributions.\label{fig:polar}}
\label{fig-1}
\end{figure*}

To explain the correspondence of multiplicity and energy distributions (schematically illustrated in Figure~\ref{fig-1}), let us consider a simple example. 
For fixed number of particles $N$, energy distribution emerges directly from the calculus of probability for a situation known as \textit{induced partition}~\cite{Fellerbook}. In short: $N-1$ randomly chosen independent points $\{U_{1},\,\ldots,\,U_{N-1}\}$ split a segment $(0,U)$ into $N$ parts, whose length is distributed according to: 
\begin{equation}
F\left(E|N\right)=\frac{N-1}{U}\left(1-\frac{E}{U}\right)^{N-2}.
\label{eq:cp}
\end{equation}
The length of the k$th$ part corresponds to the value of energy $E_{k}=U_{k+1}-U_{k}$ (for ordered $U_{k}$). Whereas for fixed $N$ one have (\ref{eq:cp}), then for $N$ fluctuating according to $P\left(N\right)$, the resulting energy distribution is 
\begin{equation}
F\left(E\right)=\sum_{N=2}^{\infty}P\left(N\right)F\left(E|N\right).
\label{eq:ed}
\end{equation}
For $P\left(N\right)$ given by BD, PD, and NBD, equation~(\ref{eq:ed}) leads to Tsallis distribution given by equation~(\ref{hag-form}).
Relationships between Poissonian multiplicity distribution and Boltzmann-Gibbs energy distribution are discussed 
in more detail in the Appendix.

Note that $P\left(N\right)$, defined for $N>1$, describe multiplicity distribution in the full phase-space. In experiments, particle multiplicity is measured usually only within some window of phase-space.  Let us assume that the detection process is a Bernoulli process described by the BD ($K=1$ and $p=\alpha$ for a fixed experimental acceptance $\alpha<1$). The number of registered particles is
\begin{equation}
M=\sum_{i=1}^{N}n_{i},
\end{equation}
where $n_{i}$ follows the BD with the generating function $G_{BD}\left(z\right)$ and $N$ comes from $P\left(N\right)$ with the generating function $G\left(z\right)$. The measured multiplicity distribution 
\begin{equation}
P\left(M\right)=\frac{1}{M!}\frac{d^{M}H\left(z\right)}{dz^{M}}\Biggl\vert_{z=0}
\end{equation}
is therefore given by generating function $H\left(z\right)=G\left(G_{BD}\left(z\right)\right)$. Such rough procedure applied to NBD, BD or PD gives again the same distributions but with modified parameters: $p\rightarrow \alpha p/\left[1-p\left(1-\alpha\right)\right]$ for NBD, $p\rightarrow \alpha p$ for BD, and $\lambda\rightarrow\alpha\lambda$ for PD. The measured multiplicity distribution is given by
\begin{equation}
P\left(M\right)=\sum_{N=M}^{\infty}P\left(N\right)P\left(M|N\right)
\end{equation}
with the acceptance function
\begin{equation}
P\left(M|N\right)=\frac{N!}{M!\left(N-M\right)!}\alpha^{M}\left(1-\alpha\right)^{N-M}
\end{equation}
Detection process extend $P\left(M\right)$ distribution to multiplicities $M=0$ and $M=1$, namely: $P\left(0\right)=\sum_{N=2}^{\infty}P\left(N\right)\left(1-\alpha\right)^{N}$ and $P\left(1\right)=\sum_{N=2}^{\infty}P\left(N\right)N\alpha\left(1-\alpha\right)^{N-1}$.

The statistical properties of the energy division between a set of particles are completely characterized by the generating function $G(z)$.
Despite correspondence between multiplicity and energy distributions, the multiplicity distribution gives in practice complementary information to the energy distribution, because $P(N)$ is defined by the $N^{th}$ derivative of $G(z)=F(E)$ at $E=U$, i.e., in the region not available experimentally in measurements at collider experiments~\footnote{Similarly as $N^{th}$ derivatives of $G(z)$ taken at $z=0$ define multiplicity distribution $P(N)$, the respective derivatives taken at $z=1$ define factorial moments $\mathcal{F}_{N}$. Derivatives of $\ln(G(z))$ taken at $z=0$ and $z=1$ define combinants $\mathcal{C}_{N}$ and cumulant factorial moments $\mathcal{K}_{N}$, respectively.}.

The above considerations (in particular equality given by equation (\ref{hag-form}) apply to a single statistical ensembles (as realized in hadronic collisions). In nuclear collisions there are usually many statistical systems, independent from one another. In superposition models of hadron production, the number of particles $N$, as registered in the experiment, is composed from independent production from $N_S$ sources~\cite{Broniowski:2017tjq}. For a fixed number of sources (neglecting the nuclear modification factor) we have $F_{AA}(E)=N_{S}\cdot F_{pp}(E)$ and $G_{AA}(z)=(G_{pp}(z))^{N_{S}}$, what results in equality:
\begin{equation}
F_{AA}(E)=N_{S}\cdot \left(G_{AA}\left(z=1-E/U_{pp}\right)\right)^{1/N_{S}}.
\end{equation}

For fluctuating numbers of sources, the resulting multiplicity distribution is given by the compound distribution defined by generating function $G_{AA}(z)=H(G_{pp}(z))$, where $H(z)$ is the generating function of distribution of the number of sources. In this case we have a relationship $F_{AA}(E)=\langle N_{S}\rangle H^{-1}\left[G_{AA}(z)\right]$, where $H^{-1}$ is the inverse function to $H(z)$, what is troublesome in practical applications.

\vspace{1.cm}

This research was supported by the Polish National Science Centre grant 2016/23/B/ST2/00692 (MR).

\appendix
\section{Boltzmann-Gibbs energy distribution and Poissonian multiplicity distribution}
\setcounter{equation}{0}

Suppose that one has $N$ independently produced particles with energies $\{E_{1,\ldots,N}\}$, distributed according to Boltzmann distribution,
\begin{equation}
F\left(E\right)=\frac{1}{T}\exp\left(-\frac{E}{T}\right)
\label{eq:BG}
\end{equation}
with ``temperature'' parameter $T=\langle E\rangle$. The sum of energies, $U=\sum_{i=1}^{N}E_{i}$ is then distributed according to gamma distribution
\begin{align}
F_{N}\left(U\right)&=\frac{1}{T\left(N-1\right)!}\left(\frac{U}{T}\right)^{N-1}\exp\left(-\frac{U}{T}\right)\nonumber\\
                   &=F_{N-1}\left(U\right)\frac{U}{N-1}
\end{align}
with cumulative distribution equal to:
\begin{equation}
F_{N}\left(>U\right)=1-\sum_{i=1}^{N-1}\frac{1}{\left(i-1\right)!}\left(\frac{U}{T}\right)^{i-1}\exp\left(-\frac{U}{T}\right).
\end{equation}

Looking for such $N$ that $\sum_{i=0}^{N}E_{i}\leq U\leq \sum_{i=0}^{N+1}E_{i}$ we find its distribution. which has known Poissonian form
\begin{align}
P\left(N\right)&=F_{N+1}\left(>U\right)-F_{N}\left(>U\right)\nonumber\\
               &=\frac{\left(U/T\right)^{N}}{N!}\exp\left(-\frac{U}{T}\right)\nonumber\\
               &=\frac{\langle N\rangle^{N}}{N!}\exp\left(-\langle N\rangle\right)
\label{eq:Pois}
\end{align}
with $\langle N\rangle=U/T$.

For the constrained systems (if the available energy is limited, $U={\rm const}$), whenever we have independent variables $\{E_{1,\ldots,N}\}$ taken from the exponential distribution (\ref{eq:BG}), the corresponding multiplicity $N$ has Poissonian distribution (\ref{eq:Pois})~\footnote{Actually this is the method of generating Poisson distribution in the numerical Monte Carlo codes.}. However, if the multiplicity is limited, $N={\rm const}$, the resulting \textit{conditional probability} becomes:
\begin{align}
F\left(E|N\right)&=\frac{F_{1}\left(E\right)F_{N-1}\left(U-E\right)}{F_{N}\left(U\right)}\nonumber\\
               &=\frac{N-1}{U}\left(1-\frac{E}{U}\right)^{N-2}
\end{align}
the same as given by equation~(\ref{hag-form}), and only in the limit $N\rightarrow \infty$ the energy distribution goes to the Boltzmann distribution (\ref{eq:BG}). For fluctuating multiplicity according to Poisson distribution, the energy distribution is given by (\ref{eq:BG}).

In the same way, as demonstrated in Ref.~\cite{Wilk:2007}, Tsallis energy distribution is connected with the NBD of multiplicity.

\vspace{-2mm}



\begin{thebibliography}{99}

%\cite{VanDantzig}
\bibitem{VanDantzig}
D. Van Dantzig, Colloques internationaux du CNRS {\bf 13} 29-45 (1949).

%\cite{Runnenburg}
\bibitem{Runnenburg}
J.T. Runnenburg, \textit{On the use of Collective Marks in Queueing Theory}. In
W.L. Smith and W.E. Wilkinson, editors, \textit{Congestion Theory} pp. 399-438.
(University of North Carolina Press, Chapel Hill, 1965).

%\cite{Kleinrock}
\bibitem{Kleinrock}
L. Kleinrock, \textit{Queueing Systems} Volume 1, Chapter 7 (Wiley, New York, 1975).

%\cite{Zhang}
\bibitem{Zhang} 
  Y.~Zhang, M.~Hlynka, P.~H.~Brill,
  %``Collective marks and first passage times,''
  arXiv:1908.04370v1 [math.PR].
  %%CITATION = ARXIV:1908.04370;%%

%\cite{Tsallis:1987eu}
\bibitem{Tsallis:1987eu} 
  C.~Tsallis,
  %``Possible Generalization of Boltzmann-Gibbs Statistics,''
  J.\ Statist.\ Phys.\  {\bf 52}, 479 (1988).
  doi:10.1007/BF01016429
  %%CITATION = doi:10.1007/BF01016429;%%
  %1104 citations counted in INSPIRE as of 08 Jan 2020

%\cite{Tsallis:2008mc}
\bibitem{Tsallis:2008mc} 
  C.~Tsallis,
  %``Nonadditive entropy: The Concept and its use,''
  Eur.\ Phys.\ J.\ A {\bf 40}, 257 (2009)
  doi:10.1140/epja/i2009-10799-0
  [arXiv:0812.4370 [physics.data-an]].
  %%CITATION = doi:10.1140/epja/i2009-10799-0;%%
  %63 citations counted in INSPIRE as of 08 Jan 2020
  
%\cite{Tsallisbook}
\bibitem{Tsallisbook} 
C. Tsallis, \textit{Introduction to Nonextensive Statistical Mechanics} (Springer, Berlin, 2009).

%\cite{Wilk:2014zka}
\bibitem{Wilk:2014zka} 
  G.~Wilk and Z.~Włodarczyk,
  %``Quasi-power law ensembles,''
  Acta Phys.\ Polon.\ B {\bf 46}, no. 6, 1103 (2015)
  doi:10.5506/APhysPolB.46.1103
  [arXiv:1501.01936 [cond-mat.stat-mech]].
  %%CITATION = doi:10.5506/APhysPolB.46.1103;%%
  %24 citations counted in INSPIRE as of 08 Jan 2020

%\cite{Michael:1976pz}
\bibitem{Michael:1976pz} 
  C.~Michael and L.~Vanryckeghem,
  %``Consequences of Momentum Conservation for Particle Production at Large Transverse Momentum,''
  J.\ Phys.\ G {\bf 3}, L151 (1977).
  doi:10.1088/0305-4616/3/8/002
  %%CITATION = doi:10.1088/0305-4616/3/8/002;%%
  %35 citations counted in INSPIRE as of 09 Jan 2020

%\cite{Michael:1977hx}
\bibitem{Michael:1977hx} 
  C.~Michael,
  %``Large Transverse Momentum and Large Mass Production in Hadronic Interactions,''
  Prog.\ Part.\ Nucl.\ Phys.\  {\bf 2}, 1 (1979).
  doi:10.1016/0146-6410(79)90002-4
  %%CITATION = doi:10.1016/0146-6410(79)90002-4;%%
  %30 citations counted in INSPIRE as of 09 Jan 2020

%\cite{Arnison:1982ed}
\bibitem{Arnison:1982ed} 
  G.~Arnison {\it et al.} [UA1 Collaboration],
  %``Transverse Momentum Spectra for Charged Particles at the CERN Proton anti-Proton Collider,''
  Phys.\ Lett.\  {\bf 118B}, 167 (1982).
  doi:10.1016/0370-2693(82)90623-2
  %%CITATION = doi:10.1016/0370-2693(82)90623-2;%%
  %454 citations counted in INSPIRE as of 09 Jan 2020

%\cite{Hagedorn:1983wk}
\bibitem{Hagedorn:1983wk} 
  R.~Hagedorn,
  %``Multiplicities, $p_T$ Distributions and the Expected Hadron $\to$ Quark - Gluon Phase Transition,''
  Riv.\ Nuovo Cim.\  {\bf 6N10}, 1 (1983).
  doi:10.1007/BF02740917
  %%CITATION = doi:10.1007/BF02740917;%%
  %106 citations counted in INSPIRE as of 09 Jan 2020

%\cite{Wong:2015mba}
\bibitem{Wong:2015mba} 
  C.~Y.~Wong, G.~Wilk, L.~J.~L.~Cirto and C.~Tsallis,
  %``From QCD-based hard-scattering to nonextensive statistical mechanical descriptions of transverse momentum spectra in high-energy $pp$ and $p\bar p$ collisions,''
  Phys.\ Rev.\ D {\bf 91}, no. 11, 114027 (2015)
  doi:10.1103/PhysRevD.91.114027
  [arXiv:1505.02022 [hep-ph]].
  %%CITATION = doi:10.1103/PhysRevD.91.114027;%%
  %54 citations counted in INSPIRE as of 09 Jan 2020

%\cite{Wilk:2012zn}
\bibitem{Wilk:2012zn} 
  G.~Wilk and Z.~Wlodarczyk,
  %``Consequences of temperature fluctuations in observables measured in high energy collisions,''
  Eur.\ Phys.\ J.\ A {\bf 48}, 161 (2012)
  doi:10.1140/epja/i2012-12161-y
  [arXiv:1203.4452 [hep-ph]].
  %%CITATION = doi:10.1140/epja/i2012-12161-y;%%
  %47 citations counted in INSPIRE as of 09 Jan 2020

%\cite{Fellerbook}
\bibitem{Fellerbook} 
W. Feller, \textit{An introduction to probability theory and its applications}, Volume II, (John Wiley and Sons Inc., New York, 1966).

%\cite{Broniowski:2017tjq}
\bibitem{Broniowski:2017tjq}
W.~Broniowski and A.~Olszewski,
%``Statistical moments in superposition models and strongly intensive measures,''
Phys. Rev. C \textbf{95} (2017) no.6, 064910
doi:10.1103/PhysRevC.95.064910
[arXiv:1704.01532 [nucl-th]].
%8 citations counted in INSPIRE as of 06 Nov 2020

%\cite{Wilk:2007}
\bibitem{Wilk:2007} 
  G.~Wilk and Z.~Wlodarczyk,
  %``Fluctuations, Correlations and the nonextensivity,''
  Physica\ A {\bf 376}, 279 (2007)

\end{thebibliography}
\end{document}